\documentclass[12pt]{article}

\usepackage{graphicx}

\newcommand{\tr}{\mbox{Tr} \, }
\newcommand{\ket}[1]{\left | #1 \right \rangle}
\newcommand{\bra}[1]{\left \langle #1 \right |}
\newcommand{\amp}[2]{\left \langle #1 \left | #2 \right. \right \rangle}
\newcommand{\proj}[1]{\ket{#1} \! \bra{#1}}

\newcommand{\superop}{{\cal E}}

\newcommand{\relent}[2]{{\cal S}\left ( #1 || #2 \right )}
\newcommand{\classrelent}[2]{{\cal H}\left ( #1 || #2 \right )}
\newcommand{\supp}{\mbox{supp} \, }

\begin{document}

\title{Approximate quantum error correction}
\author{Benjamin Schumacher$^{(1)}$
        and Michael D. Westmoreland$^{(2)}$}
\maketitle
\begin{center}
{\sl
$^{(1)}$Department of Physics, Kenyon College, Gambier, OH 43022 USA \\
$^{(2)}$Department of Mathematical Sciences, Denison University,
 Granville, OH  43023 USA }
\end{center}

\section*{Abstract}

The errors that arise in a quantum channel can be corrected perfectly
if and only if the channel does not decrease the coherent information
of the input state.  We show that, if the loss of coherent information
is small, then {\em approximate} quantum error correction is possible.

\section{Perfect quantum error correction}

The problem of quantum information transfer via a channel
can be cast as the problem of sending quantum entanglement using
the channel \cite{s96}.  Suppose $R$ and $Q$ are subsystems of a composite
quantum system initially in a pure entangled state $\ket{\Psi^{RQ}}$.
This state has a Schmidt decomposition
\begin{equation}
	\ket{\Psi^{RQ}} = \sum_{k} \sqrt{\lambda_k} \ket{k^R} \otimes \ket{k^Q}
\end{equation}
where the $\{ \ket{k^R} \}$ and $\{ \ket{k^Q} \}$ are orthonormal sets
of $R$ and $Q$ states, respectively.

System $Q$ is transmitted from the sender to the receiver via
a noisy channel while $R$ remains isolated at the sender's end.
In the noisy case, the evolution of $Q$ involves a unitary
interaction with an environment system $E$ (initially in some
state $\ket{0^E}$), leading to a final joint state
\begin{equation}
	\ket{\Psi^{RQE'}} =  \left ( 1^{R} \otimes U^{QE} \right )
				\ket{\Psi^{RQ}} \otimes \ket{0^{E}} .
\end{equation}
The (mixed) states of various subsystems are obtained by partial traces.
The net effect of the interaction of $Q$ with $E$ is that the evolution of
$Q$ by itself is described by a superoperator $\superop^{Q}$.

An error correction procedure is an operation that is performed
by the receiver (and thus on $Q$) with the aim of restoring the
output state $\rho^{RQ'}$ of $RQ$ to the input state $\ket{\Psi^{RQ}}$.
Of course, this is not always possible.  Schumacher and Nielsen
\cite{sn96} provided a necessary and sufficient condition for
the existence of a perfect error correction procedure---that is,
one that exactly restores the input state.  This condition is
based on the coherent information of the process.

\begin{figure}
\unitlength 0.75cm
\begin{center}
\begin{picture}(8,7)(0,0)
% Q --> Q'
\put(2,3){\framebox(1,1){$Q$}}
\put(3,3.5){\vector(1,0){3}}
\put(6,3){\framebox(1,1){$Q'$}}
%
% add R plus initial RQ state
\put(2,5){\framebox(1,1){$R$}}
\put(2.4,4.10){\line(0,1){.05}}	\put(2.6,4.10){\line(0,1){.05}}
\put(2.4,4.25){\line(0,1){.05}}  \put(2.6,4.25){\line(0,1){.05}}
\put(2.4,4.40){\line(0,1){.05}}  \put(2.6,4.40){\line(0,1){.05}}
\put(2.4,4.55){\line(0,1){.05}}  \put(2.6,4.55){\line(0,1){.05}}
\put(2.4,4.70){\line(0,1){.05}}  \put(2.6,4.70){\line(0,1){.05}}
\put(2.4,4.85){\line(0,1){.05}}  \put(2.6,4.85){\line(0,1){.05}}
\put(1,4.5){\makebox(0,0){$\ket{\Psi^{RQ}}$}}
%
% purify first stage dynamics
\put(4,0.5){\framebox(1,1){$E$}}
\put(4.4,1.7){\vector(0,1){1.6}}
\put(4.6,3.3){\vector(0,-1){1.6}}
\put(4.8,2.5){\makebox(0,0)[cl]{$U^{QE}$}}
\end{picture}
\end{center}
\caption{Composite system $RQ$ is initially in the entangled state
       $\ket{\Psi^{RQ}}$.  The evolution of system $Q$ includes
       interaction with the environment $E$.}
\end{figure}
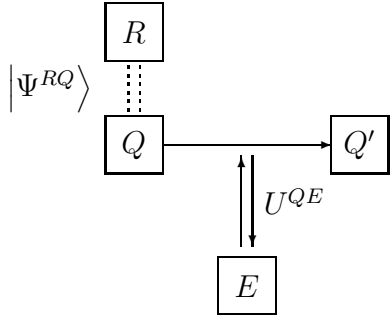

The coherent information $I$ is defined to be
\begin{equation}
	I = S^{Q'} - S^{RQ'}
\end{equation}
where $S^{Q'}$ and $S^{RQ'}$ are the von Neumann entropies of the systems
$Q$ and $RQ$ after the evolution:  $S(\rho) = -\tr \rho \log \rho$.
The coherent information $I \leq S^Q$
(the initial entropy of entanglement between $R$ and $Q$), since
\begin{equation}
	S^Q - I = S^Q + S^{RQ'} - S^{Q'} = S^{R'} + S^{E'} - S^{RE'} \geq 0
	\label{decoherence}
\end{equation}
by the subadditivity of the entropy.  (We have also used the fact that the
overall state of $RQE$ is always a pure state, and that the ``reference''
system $R$ never interacts with anything.)

Perfect quantum error correction means that an operation on $Q$ alone can
restore the joint state of $RQ$ to the original state $\ket{\Psi^{RQ}}$ with
fidelity equal to unity.  In \cite{sn96} it is shown that perfect quantum
error correction is possible if and only if $I = S^Q$.

``Only if'' follows from the fact that the coherent information $I$ cannot
be increased by any subsequent quantum data processing; thus, any non-zero
loss of coherent information (that is, any positive decoherence $D$) can
never be restored.  The ``if'' part is more interesting.  By Equation~\ref{decoherence}
we see that $I = S^Q$ implies that $S^{R'} + S^{E'} = S^{RE'}$.
This is only true when $RE$ is in a product state:
\begin{equation}
	\rho^{RE'} = \rho^{R'} \otimes \rho^{E'} .
\end{equation}
Since the overall final state of $RQE$ is pure, we can write it as
\begin{equation}
	\ket{\Psi^{RQE'}} = \sum_{kl} \sqrt{\lambda_k \mu_l}
            \ket{k^R} \otimes \ket{\phi_{kl}^{Q}} \otimes \ket{l^E} .
	\label{finalprodstate}
\end{equation}
Here the $\lambda_k$ are the eigenvalues of $\rho^R$ (unchanged by the
dynamics) and the $\mu_l$ and $\ket{l^E}$ are the eigenvalues and
eigenstates of $\rho^{E'}$.  The states $\ket{\phi_{kl}^Q}$ are orthonormal.

We can exploit the structure of $\ket{\Psi^{RQE'}}$ given in
Equation~\ref{finalprodstate} to construct an error-correction
procedure for $Q$.  First, we make an incomplete ideal projective
measurement on $Q$, where the projection operators are given by
\begin{equation}
	\Pi_l = \sum_{k} \proj{\phi_{kl}^Q} .
\end{equation}
The outcome $l$ of this measurement on the state $\ket{\Psi^{RQE'}}$ will
occur with probability $\mu_l$, and the resulting state will be
\begin{equation}
	\ket{\psi^{RQE'}_l} = \sum_{k} \sqrt{\lambda_k} \ket{k^R} \otimes
               \ket{\phi_{kl}^Q} \otimes \ket{l^E} .
\end{equation}
We now make a unitary transformation of $Q$ that is conditional on the
measurement outcome.  If the outcome is $l$, we apply a unitary operator
$U_l$ such that
\begin{equation}
	U_l \ket{\phi_{kl}^Q} = \ket{k^Q} .
\end{equation}
After this transformation, the final state is
\begin{equation}
	U_l \ket{\psi^{RQE'}_l} = \left ( \sum_k \sqrt{\lambda_k} \ket{k^R} \otimes
              \ket{k^Q} \right ) \otimes \ket{l^E}
\end{equation}
and so the state of $RQ$ has been perfectly restored to $\ket{\Psi^{RQ}}$.

If we can perfectly restore $\ket{\Psi^{RQ}}$ by this procedure, then the same
procedure will allow us to restore any pure state $\ket{\phi^Q}$ in the 
support of $\rho^Q = \tr_R \proj{\Psi^{RQ}}$.  The converse is also true;
that is, the problem of faithfully transmitting the entanglement between $R$ and $Q$
is equivalent to the problem of sending an arbitrary pure state of $Q$ through
the channel.

We now turn to our main question.  Suppose $I < S^Q$, but the difference between
them is small.  Perfect quantum error correction is not possible, but it seems
plausible that {\em nearly} perfect error correction is.  With what fidelity
can we restore the original state $\ket{\Psi^{RQ}}$ by an error correction
procedure on $Q$?  To answer this, we will first describe the connections between
various measures of the ``closeness'' of two quantum states, with the goal of
forging a link between an entropic measure of closeness (related to the coherent
information) and fidelity.

\section{Distance measures}

The relative entropy function $\relent{\rho}{\sigma}$ between two
density operators $\rho$ and $\sigma$ is defined to be
\begin{equation}
	\relent{\rho}{\sigma} = \tr \rho \log \rho - \tr \rho \log \sigma.
\end{equation}
The relative entropy function is not a metric, but
it does have some of the intuitive properties of a ``distance'' between
density operators.  For example, the relative entropy is never negative,
and is zero if and only if $\rho = \sigma$.

The trace distance $D(\rho,\sigma)$ is a metric on the set of density
operators.  It is defined by
\begin{equation}
	D(\rho,\sigma) = \frac{1}{2} \, \tr | \rho - \sigma |,
\end{equation}
where the factor of one-half is chosen to agree with the definition in
\cite{mikeandike}.

If $\rho$ is a density operator, then $\ket{\psi_\rho}$ represents a
{\em purification} of $\rho$--that is, a pure state of a larger system which
reduces to the state $\rho$ under partial trace.  A given density operator
$\rho$ has many purifications $\ket{\psi_\rho}$.
The fidelity $F(\rho, \sigma)$ between two density operators is
\begin{equation}
	F(\rho, \sigma) = \max \left | \amp{\psi_\rho}{\psi_\sigma} \right |
	\label{fidelitydef}
\end{equation}
where the maximum is taken over all choices of purifications $\ket{\psi_\rho}$
and $\ket{\psi_\sigma}$.  (Equivalently, we may fix one of the purifications
$\ket{\psi_\rho}$ and maximize only over the other one.)  The fidelity
satisfies
\begin{equation}
	0 \leq F(\rho,\sigma) \leq 1 ,
\end{equation}
with the fidelity equal to unity if and only if $\rho = \sigma$.
Relative entropy, trace distance, and fidelity all provide measures of the
``closeness'' of two density operators.

Suppose $\superop$ is a generalized quantum operation---that is,
a trace-preserving completely positive map on density operators.
Then all three notions of ``closeness'' are monotonic,
in the following sense:
\begin{eqnarray}
	\relent{\superop(\rho)}{\superop(\sigma)} & \leq & \relent{\rho}{\sigma}
					\nonumber \\
	D(\superop(\rho),\superop(\sigma)) & \leq & D(\rho,\sigma) \\
	F(\superop(\rho),\superop(\sigma)) & \geq & F(\rho,\sigma) . \nonumber
\end{eqnarray}

In \cite{mikeandike}, it is shown that the trace distance is related
to the fidelity $F(\rho,\sigma)$ between the two density operators:
\begin{equation}
	F(\rho,\sigma) \geq 1 - D(\rho,\sigma) .
\end{equation}
Thus, two density operators that are ``close'' in trace distance are also
``close'' in fidelity.

Now we will show that relative entropy and trace distance are related by
\begin{equation}
	\relent{\rho}{\sigma} \geq \frac{2}{\ln 2} \,
                          \left ( D(\rho,\sigma) \right )^2 .
	\label{distrelation}
\end{equation}
To prove Equation~\ref{distrelation}, first suppose that we
perform a fixed measurement on a system that may be either
in state $\rho$ or state $\sigma$,
leading to probability distributions $\vec{P}$ and $\vec{Q}$,
respectively.  Then the classical relative entropy of the
probability distributions is no greater than the quantum relative
entropy of the density operators:
\begin{equation}
	\relent{\rho}{\sigma} \geq \classrelent{\vec{P}}{\vec{Q}}
	\label{qtocrelent}
\end{equation}
where $\classrelent{\vec{P}}{\vec{Q}}$ is the classical relative entropy
function
\begin{equation}
	\classrelent{\vec{P}}{\vec{Q}} = \sum_{k} P_k \log P_k \, - \,
		\sum_{k} P_k \log Q_k  .
\end{equation}
Equation~\ref{qtocrelent} is true be cause the measurement process,
leading to classically distinguishable states of the measuring apparatus,
is a generalized quantum operation.  The relative entropy cannot
increase in this process.
In \cite{cover} it is shown that, given two probability distributions
$\vec{P}$ and $\vec{Q}$,
\begin{equation}
	\classrelent{\vec{P}}{\vec{Q}} \geq \frac{1}{2 \ln 2}
			\left ( \sum_{k} \left | P_k - Q_k \right | \right )^2 .
	\label{classdistrelation}
\end{equation}

Next, consider the operator $\rho - \sigma$.  This operator can be written
\begin{displaymath}
	\rho - \sigma = A - B
\end{displaymath}
where $A$ and $B$ are positive operators such that $\supp A$ and $\supp B$ are
orthogonal subspaces.  Notice that $\tr A = \tr B$.
Let $\Pi$ be the projection onto the support of $A$.
Then $| \rho - \sigma | = A + B$ and
\begin{eqnarray*}
	D(\rho,\sigma) & = & \frac{1}{2} \, \tr | \rho - \sigma | \\
			& = & \frac{1}{2} \left ( \tr A + \tr B \right ) \\
			& = & \tr A \\
			& = & \tr \Pi (\rho - \sigma) .
\end{eqnarray*}
Suppose we perform the projective measurement corresponding to $\Pi$ and $1-\Pi$.
If our state is $\rho$, we obtain probabilities
$P_1 = \tr \Pi \rho$ and $P_2 = 1 - P_1$.  If our state is $\sigma$, we obtain
$Q_1 = \tr \Pi \sigma$ and $Q_2 = 1 - Q_2$.  Then by the definition of $\Pi$,
\begin{eqnarray*}
	P_1 - Q_1 & = & \tr \Pi (\rho - \sigma) \\
		  & \geq & 0 \\
	P_2 - Q_2 & = & \tr (1 - \Pi) (\rho - \sigma) \\
		  & = & - \tr \Pi (\rho - \sigma)
\end{eqnarray*}
and thus
\begin{equation}
	\sum_{k} \left | P_k - Q_k \right | = 2 \, \tr \Pi (\rho - \sigma)
		= 2 \, D(\rho,\sigma) .
\end{equation}
Combining this with Equations~\ref{qtocrelent} and \ref{classdistrelation} we
obtain
\begin{displaymath}
	\relent{\rho}{\sigma} \geq \frac{2}{\ln 2} \,
                          \left ( D(\rho,\sigma) \right )^2
\end{displaymath}
which is Equation~\ref{distrelation}, as desired.

In terms of fidelity, we have
\begin{displaymath}
	F(\rho,\sigma) \geq 1 - \sqrt{\frac{\ln2}{2} \, \relent{\rho}{\sigma}}
			\geq 1 - \sqrt{\relent{\rho}{\sigma}} .
\end{displaymath}
Thus, if $\relent{\rho}{\sigma} < \epsilon$, then $F(\rho,\sigma) >
1 - \sqrt{\epsilon}$.
States that are close in the relative entropy sense are also close in fidelity.

\section{Approximate error correction}

We now return to our discussion of approximate error correction
in a quantum channel in which the coherent information $I$ is
only slightly less than $S^Q$, the initial entropy of entanglement
of $RQ$.
To make the question precise, suppose that $S^Q - I < \epsilon$.
With what fidelity can we restore the initial
state $\ket{\Psi^{RQ}}$ by an operation on $Q$?

We first note that
\begin{eqnarray}
S^{Q} - I & = & S^{R'} + S^{E'} - S^{RE'} \nonumber \\
          & = & \relent{\rho^{RE'}}{\rho^{R'} \otimes \rho^{E'}} .
\end{eqnarray}
Thus our hypothesis reduces to
\begin{displaymath}
\relent{\rho^{RE'}}{\rho^{R'} \otimes \rho^{E'}} < \epsilon .
\end{displaymath}
The state $\rho^{RE'}$ is not a product state, but it is close (in the
relative entropy sense) to the product state $\rho^{R'} \otimes \rho^{E'}$.
We can therefore assert that the fidelity
\begin{displaymath}
	F(\rho^{RE'},\rho^{R'} \otimes \rho^{E'}) > 1 - \sqrt{\epsilon} .
\end{displaymath}

The actual final state $\ket{\Psi^{RQE'}}$ of the overall system $RQE$ is
a purification of $\rho^{RE'}$.  Denote by $\ket{\hat{\Psi}^{RQE'}}$ a
specific purification of $\rho^{R'} \otimes \rho^{E'}$ for which
\begin{equation}
	F(\rho^{RE'},\rho^{R'} \otimes \rho^{E'}) =
		\left | \amp{\Psi^{RQE'}}{\hat{\Psi}^{RQE'}} \right | .
\end{equation}
(Such a purification must, of course, exist by our definition of the
fidelity.)  Since $\rho^{R'} \otimes \rho^{E'}$ is a product state, this
purification has the form
\begin{equation}
	\ket{\hat{\Psi}^{RQE'}} = \sum_{kl} \sqrt{\lambda_k \mu_l}
            \ket{k^R} \otimes \ket{\phi_{kl}^{Q}} \otimes \ket{l^E} ,
\end{equation}
Because this state has exactly the same form as the one in
Equation~\ref{finalprodstate}, we can as before construct an error
correction operation on $Q$ that would exactly restore the state
$\ket{\hat{\Psi}^{RQE'}}$ to the original entangled state
$\ket{\Psi^{RQ}}$ on $RQ$.  Denote this operation by $\superop^Q$.

We do not actually have the product state purification $\ket{\hat{\Psi}^{RQE'}}$.
Instead, we have the nearby final state $\ket{\Psi^{RQE'}}$.  However, we know
that the fidelity between two states cannot be decreased by any operation.
Therefore, if we apply the operation $\superop^Q$ to the actual final state,
we'll arrive at an error corrected state $\omega^{RQ}$ for $RQ$, such that
\begin{eqnarray}
	F(\omega^{RQ},\ket{\Psi^{RQ}})
		& \geq & \left | \amp{\Psi^{RQE'}}{\hat{\Psi}^{RQE'}} \right |
					\nonumber \\
		& = & F(\rho^{RE'},\rho^{R'} \otimes \rho^{E'}) \nonumber \\
		& > & 1 - \sqrt{\epsilon} .
\end{eqnarray}
Therefore, if the coherent information decreases by less than $\epsilon$, we
can find an error correction procedure that restores the original entangled
state with a fidelity greater than $1 - \sqrt{\epsilon}$.

The {\em entanglement fidelity} $F_e$ of the channel (defined in \cite{s96})
is related to the fidelity discussed here by $F_e = F^2$.  Thus, we have shown
that it is possible to correct errors with entanglement fidelity
\begin{equation}
	F_e > \left ( 1 - \sqrt{\epsilon} \right )^2 \geq 1 - 2 \sqrt{\epsilon} .
\end{equation}

We now see that the necessary and sufficient criterion for
perfect quantum error correction derived in \cite{sn96} is
{\em robust}.  If it is approximately satisfied, then
errors can be approximately corrected.  This result is likely
to be more useful in analyzing the capacities of a quantum channel,
since in general we would not require that the fidelity ever
exactly equals unity, but only that it approaches unity
asymptotically.

An approximate quantum error correction scheme has been presented
by Barnum and Knill \cite{barnumknill}, though not in the context
of small loss of coherent information.  The relationship, if any,
between their scheme and the one presented here is not known.

We are very grateful to M. A. Nielsen for comments and suggestions.

\end{document}